\documentstyle[prd,aps,preprint,tighten,eqsecnum,epsf]{revtex}
\begin{document}
\newcommand{\naught}{{\scriptscriptstyle 0}}
\newcommand{\ed}{\mbox{$\partial \hspace{-2mm} /$}}
\newcommand{\ednaught}{\ed{}_{\!\naught}}
\newcommand{\edbarnaught}{\bar{\ed}{}_{\!\naught}}
\preprint{gr-qc/9609057}

\title{Energy of Isolated Systems at Retarded Times as the\\ 
Null Limit of Quasilocal Energy\footnote{Preprint numbers:
NCSU/CTMP/013, TUW-10-96, IFP-730-UNC and TAR-054-UNC}} 

\author{J. David Brown\footnote{Email 
addresses: david\_brown@ncsu.edu,
lau@tph16.tuwien.ac.at, and york@physics.unc.edu}}
\address{Department of Physics and Department of Mathematics,\\
North Carolina State University,\\
Raleigh, NC 27695-8202 USA}
 
\author{Stephen R. Lau}
\address{Institut f\"{u}r Theoretische Physik,\\
Technische Universit\"{a}t Wien,\\
Wiedner Hauptstra\ss e 8-10,\\
A-1040 Wien, \"{O}sterreich}

\author{James W. York, Jr.}
\address{Institute of Field Physics and\\ 
Theoretical Astrophysics and Relativity Group,\\
Department of Physics and Astronomy,\\
University of North Carolina,\\
Chapel Hill NC, 27599-3255 USA}
\maketitle
\begin{abstract}
We define the energy of a perfectly isolated system at a given 
retarded time as the suitable null limit of the quasilocal energy 
$E$. The result coincides with the Bondi-Sachs mass. Our $E$ is the 
lapse-unity shift-zero boundary value of the gravitational 
Hamiltonian appropriate for the partial system $\Sigma$ contained 
within a finite topologically spherical boundary $B = \partial \Sigma$. 
Moreover, we show that with an arbitrary lapse and zero shift the 
same null limit of the Hamiltonian defines a physically 
meaningful element in the space dual to supertranslations. This 
result is specialized to yield an expression for the full 
Bondi-Sachs four-momentum in terms of Hamiltonian values.
\end{abstract}
\newpage

\section*{Introduction} 
To define quasilocal energy in general relativity, one can begin with 
a suitable action functional for the time history ${\cal M}$ of a 
spatially bounded system $\Sigma$. Here ``suitable'' means that in the 
associated variational principle the induced metric on the 
time history ${\cal T}$ of the system boundary $B = 
\partial\Sigma$ is fixed. In particular, this means that the 
lapse of proper time between the boundaries of the initial and final 
states of the system $\Sigma$ must be fixed as boundary data. 
The quasilocal energy (QLE) is then defined as minus 
the rate of change of the classical action (or Hamilton-Jacobi 
principal function) corresponding to a unit increase in proper 
time.\cite{BY,Lau} So defined, the QLE is a functional on the gravitational 
phase space of $\Sigma$, and is the value of the gravitational Hamiltonian 
corresponding to unit lapse function and zero shift vector on the 
system boundary $B$. Although other definitions of quasilocal energy 
have been proposed (see, for example, the references listed in \cite{BY}), 
the QLE considered here has the key property, which we consider crucial, 
that it plays the role of internal energy in the thermodynamical 
description of coupled gravitational and matter fields.\cite{thermo}  

In this paper we define the energy of a perfectly isolated system at a 
given retarded time as the suitable limit of the quasilocal energy $E$ 
for the partial system enclosed within a finite topologically spherical 
boundary.\footnote{Hecht and Nester have also considered energy-momentum
(and ``spin'') at null infinity (for a class of generally covariant 
theories including general relativity) {\em via} limits of quasilocal Hamiltonian 
values.\cite{HechtNester} Their treatment of energy-momentum is based on a 
differential-forms version of canonical gravity, often referred to as the 
``covariant canonical formalism." For pure {\sc bms} translations our results 
are in accord with those found by Hecht and Nester, although at the level of 
{\em general} supertranslations they differ. We provide a careful analysis 
of the zero-energy reference term (necessary for the QLE to have a finite 
limit at null infinity), and this analysis is intimately connected with 
our results concerning general supertranslations.} For our choice of 
asymptotic reference frame the energy 
that we compute equals what is usually called 
the Bondi-Sachs mass.\cite{Goldberg,review} As we shall see, our asymptotic 
reference frame defines precisely that infinitesimal generator of the 
Bondi-Metzner-Sachs ({\sc bms}) group corresponding to a pure time translation.\cite{Sachs,review,PenroseRindler} We also show that in the 
same null limit the lapse-arbitrary, shift-zero 
Hamiltonian boundary value defines a physically meaningful 
element in the space dual to 
supertranslations. This dual space element, it turns out, coincides with 
the ``supermomentum" discussed by Geroch.\cite{Geroch} Our results are 
then specialized to an expression for the full Bondi-Sachs four-momentum 
in terms of Hamiltonian values. It is already known that when
$B$ is the two-sphere at spacelike infinity, the quasilocal 
and Arnowitt-Deser-Misner\cite{ADM} notions of energy-momentum 
agree.\cite{BY,thermo} Our results therefore 
indicate that the quasilocal formalism provides a unified
Hamiltonian framework 
for describing the standard notions of gravitational 
energy-momentum in asymptopia. 

Before turning to the 
technical details, let us first present a short overview 
of our approach. Consider a spacetime ${\cal M}$ which is 
asymptotically flat at future null infinity ${\cal I}^{+}$
and a system $(w,R,\theta,\phi)$ of Bondi coordinates 
thereon.\cite{review} The retarded time $w$ labels a 
one-parameter family of outgoing null hypersurfaces 
${\cal N}(w)$. The coordinate $R$ is a luminosity 
parameter (areal radius) along the outgoing null-geodesic 
generators of the hypersurfaces ${\cal N}(w)$. 
The Bondi coordinate system
also defines a two-parameter family of topologically spherical 
two-surfaces $B(w,R)$. It suits our purposes to consider 
only a single null hypersurface of the family ${\cal N}(w)$, 
say ${\cal N}(w_{*})$, the one determined by setting $w$ equal 
to an {\em arbitrary} constant $w_{*}$. The collection 
$B(w_{*},R)$ of two-surfaces foliates ${\cal N}(w_{*})$, and
in the $R\to\infty$ limit these two-surfaces converge on an
infinite-radius {\em round} sphere $B(w_{*},\infty)$.
To streamline the presentation, we refer to our generic 
null hypersurface simply as $\cal N$; and we use the 
plain letter $B$ to denote both the $\cal N$-foliating 
collection $B(w_{*},R)$ and a single generic two-surface of 
this collection. Now, should we desire a more general 
${\cal N}$-foliating collection of two-surfaces, we could, of course, 
introduce a new radial coordinate $\bar{R}$. For a
fixed retarded time $w = w_{*}$ the new two-surfaces would 
then arise as level surfaces of constant $\bar{R}$. However, 
we shall not consider such a new radial coordinate, because 
the new two-surfaces would not necessarily converge towards a round 
sphere in the asymptotic limit. At any rate, we could handle such an 
additional kinematical freedom, were it present, by assuming that 
along each outgoing null ray $\bar{R}$ approached $R$ at a 
sufficiently fast rate in the asymptotic limit.

 Our first goal is to compute the QLE within a 
two-surface $B$ in the limit as $B$ approaches a spherical 
cut of ${\cal I}^{+}$ along the null surface $\cal N$, and to 
show that this result coincides with the Bondi-Sachs mass: 
\begin{equation}
M_{_{BS}}(w_{*}) = \lim_{R \rightarrow \infty} 
\int_{B(w_{*},R)} 
{\rm d}^{2}x \sqrt{\sigma} \varepsilon\, .
\label{limit} 
\end{equation}
Here $\varepsilon 
= (k - k |^{\scriptscriptstyle {\rm ref}})/\kappa$ 
is the quasilocal energy 
surface density with $\kappa = 8\pi$ (in geometrical units) 
and $\sigma$ is the determinant of the induced metric on $B$. 
Recall that $k$ denotes the mean curvature of 
$B$ as embedded in some {\em spacelike} spanning three-surface 
$\Sigma$. Since both $B$ and $\Sigma$ are embedded in the 
{\em physical} spacetime ${\cal M}$, we sometimes use the notation 
$\varepsilon |^{\scriptscriptstyle {\rm phy}} = k/\kappa$. Also recall that
$k |^{\scriptscriptstyle {\rm ref}}$ denotes the mean curvature 
of a surface which is isometric to $B$ but 
which is embedded in a three-dimensional {\em reference} space
different than $\Sigma$. Here we choose the reference space to be
flat Euclidean space $E^{3}$, i.\ e.\ 
we assign a flat three-slice of Minkowski spacetime the zero value of
energy.\cite{BY} Although a definition of the zero-energy reference in 
terms of flat space is neither always essential nor 
appropriate\cite{Brown}, it is the correct choice for the analysis 
of this paper. 

In order to define $k$, 
we must select such a three-surface spanning $B$ for each $R$ 
value. (For a single $B$ many different spanning 
three-surfaces will determine 
the same $k$. In fact, $k$ is determined solely by $B$ and a timelike  
unit vector field $u^\mu$ on $B$, which can be considered as the unit 
normal of a slice $\Sigma$. Thus, the continuation of $\Sigma$ 
away from $B$ is not needed; moreover, such a continuation of $\Sigma$ might 
not be defined throughout the interior of ${\cal M}$. Therefore, though 
we speak of choosing a $\Sigma$ three-surface to span $B$ 
for each $R$ value, we are really fixing only a timelike unit normal 
vector field at $B$.) For generality, we leave the choice of 
spanning three-surface $\Sigma$ essentially arbitrary at 
each $R$ value, but we do enforce a definite choice 
asymptotically. Heuristically, as $R \rightarrow \infty$ 
the $\Sigma$ three-surface spanning $B$ approaches an 
asymptotic three-surface $\Sigma_{\infty}$ which spans a
round infinite-radius spherical cut of ${\cal I}^{+}$ (see the
figure). Our construction is, as expected, sensitive to the choice of 
asymptotic three-surface $\Sigma_{\infty}$. Said another way, the 
QLE depends on the fleet of observers at $B$ whose four-velocities 
are orthogonal to the spanning three-surface at $B$. Therefore, 
one expects {\em a priori} the expression on the right-hand side 
of (\ref{limit}) to depend on the choice of asymptotic fleet 
associated with the two-sphere at 
${\cal I}^{+}$. The asymptotic fleet we choose 
corresponds to a pure {\sc bms} time translation:
each member of the asymptotic fleet rides along 
$\partial/\partial w$. Note that, although $\partial/\partial w$ 
is everywhere timelike in ${\cal M}$ (at least in the relevant exterior
regions), the {\em extension} of 
$\partial/\partial w$ to ${\cal I}^{+}$ in a conformal completion 
$\hat{\cal M}$ of the physical spacetime ${\cal M}$ is in fact a 
null vector which lies in ${\cal I}^{+}$. (While we occasionally 
find it clarifying to make reference to the concept of 
a conformal completion, we 
do not explicitly use conformal completions in this paper.) 
Therefore, heuristically, one should envision $\Sigma_{\infty}$
as a spacelike slice which becomes null asymptotically 
(see the figure).

This paper is organized as follows. In a preliminary
section we write down the familiar Bondi-Sachs 
form\cite{Sachs,Chrusciel_et_al} 
of the spacetime metric as well as asymptotic expansions 
for the associated metric coefficients.
We also introduce on $\cal M$ two 
future-pointing null vector fields $k^{\mu}$ and $l^{\mu}$ 
(do not confuse $k^{\mu}$ with the mean curvature
$k$). Both vector fields point everywhere normal to 
our collection $B(w,R)$ of two-surfaces, and 
$k_{\mu} l^{\mu} = - 1$.
Next, we construct on ${\cal N}$ a timelike vector field 
$u^{\mu} := \frac{1}{2} k^{\mu} + l^{\mu}$ (equality restricted
to ${\cal N}$), which in 
our analysis will {\em define} for each $B$ along $\cal N$ 
a spacelike spanning three-surface $\Sigma$. In $\S$II we use the $\Sigma$ 
three-surfaces determined by $u^{\mu}$ to define an unreferenced energy 
surface density $\varepsilon |^{\scriptscriptstyle {\rm phy}} = k/\kappa$
for each $B$ slice of $\cal N$ and then examine the asymptotic 
limit of $k/\kappa$. 
In $\S$III we consider the asymptotic expression for the flat-space 
reference density 
$\varepsilon |^{\scriptscriptstyle {\rm ref}} = 
k |^{\scriptscriptstyle {\rm ref}}/\kappa$, but give the
derivation of this expression in the Appendix. Next, we assemble 
the results of the previous two sections and prove the main 
claim (\ref{limit}). In $\S$IV we examine the ``smeared energy 
surface density,'' which is the Hamiltonian value corresponding 
to an arbitrary supertranslation. We then specialize our
result for the smeared energy surface density to express the 
full Bondi-Sachs four-momentum in terms of Hamiltonian values. In $\S$V we 
examine the smeared energy surface density 
{\em via} the spin-coefficient formalism, and show that it equals the 
``supermomentum'' of Geroch\cite{Geroch} as written by Dray and 
Streubel.\cite{DrayStreubel} The Appendix is 
devoted to a detailed analysis of
the reference term.
 
\section{Preliminaries}
In terms of a Bondi coordinate system the metric of our asymptotically 
flat spacetime ${\cal M}$ takes the standard 
form\cite{Sachs,Chrusciel_et_al}
\begin{equation}
g_{\mu\nu} {\rm d}x^{\mu} {\rm d}x^{\nu} = 
- UV {\rm d}w^{2} 
- 2 U {\rm d}w {\rm d}R 
+ \sigma_{ab}
  ({\rm d}x^{a} + W^{a}{\rm d}w) ({\rm d}x^{b} 
+ W^{b}{\rm d}w){\,} , 
\label{spacetimemetric}
\end{equation}
where $a,b$ are $B$ indices running over $\theta,\phi$. We assume the 
following expansions for the various metric coefficients 
above:\footnote{Up to the $\Delta$ remainder terms, our expansions in the
radial coordinate $R$ coincide with those given by Sachs; 
however, we do not assume that the $\Delta$ remainder terms are 
necessarily expandable in powers of inverse $R$, an assumption 
which would be tantamount to what Sachs calls the 
``outgoing radiation condition.''\cite{Sachs} Recently, Chru\'{s}ciel 
{\em et al.}~have shown that ``polyhomogeneous'' 
expansions in terms of $R^{-i}\log^{j}\!R$ also provide a consistent 
framework for solving the characteristic initial value problem of the 
Bondi-Sachs type.\cite{Chrusciel_et_al} They argue that the 
so-called outgoing radiation condition is overly restrictive.}
\begin{eqnarray}
V & = & 
  1 
- 2{\rm m} R^{-1} 
+ \Delta_{V}
\label{Sachs} \eqnum{\ref{Sachs}a} \\
U & = & 
  1 
- {\textstyle \frac{1}{2}}(X^{2} + Y^{2}) 
  R^{-2} + \Delta_{U}
\eqnum{\ref{Sachs}b}\\
W^{\theta} & = & 
  (2 X\cot \theta  + X_{,\theta} 
+ Y_{,\phi} \csc\theta)R^{-2} 
+ \Delta_{W^{\theta}}
\nonumber \\
W^{\phi} & = & 
  \csc\theta\left(2 Y \cot\theta  
+ Y_{,\theta} - X_{,\phi}\csc\theta\right)R^{-2} 
+ \Delta_{W^{\phi}}
\eqnum{\ref{Sachs}c} \\
\sigma_{ab} & = &
  R^{2}\delta_{ab} 
+ \left[(2 X)\theta_{,a}\theta_{,b}
+ (4 Y \sin\theta)\theta_{,(a} \phi_{,b)} 
- (2 X \sin^{2}\theta) \phi_{,a} \phi_{,b}\right]R 
+ \Delta_{\sigma_{ab}}{\,} . \eqnum{\ref{Sachs}d}
 \addtocounter{equation}{1}
\end{eqnarray}
Here $X(w,\theta,\phi)$ and $Y(w,\theta,\phi)$
are respectively the real and imaginary parts of the asymptotic 
shear $c = X + {\rm i} Y$, ${\rm m}(w,\theta,\phi)$ is the 
all-important {\em mass aspect}, $\delta_{ab}$ is the metric of a 
unit-radius round sphere, and commas denote partial differentiation. 
In the Appendix we examine the form of the two metric $\sigma_{ab}$ in 
more detail. Remainder terms, denoted by the $\Delta$ symbol, always fall 
off faster (or have slower growth, as the case may be) than the terms 
which precede them. For instance, $\Delta_{V}$ denotes a term 
which falls off {\em faster} than $O(R^{-1})$. 

Introduce the future-directed null covector field 
$k_{\mu} = - e^{\eta}
\nabla_{\mu} w$, where the scalar function
$\eta = \eta(w, R , \theta , \phi)$ is a point-dependent 
boost parameter. The null covector $k_{\mu}$ is orthogonal 
to the spheres $B(w,R)$, and the function $\eta$ gives us 
complete freedom in choosing the extent of $k_{\mu}$ at each 
point of any $B$ two-surface. We shall find it necessary later to assume that 
$\eta$ falls off {\em faster} than $1/\sqrt{R}$ on every outgoing ray. 
Also define another future-directed null 
vector field $l^{\mu}$ which is 
orthogonal to the $B(w,R)$ and normalized so that 
$k_{\mu} l^{\mu} = - 1$. As one-forms these 
null normals are 
\begin{eqnarray}
k_{\mu} {\rm d}x^{\mu} & = &
- e^{\eta} {\rm d}w
\label{bassae} \eqnum{\ref{bassae}a} \\
l_{\mu} {\rm d}x^{\mu} & = &
- e^{-\eta} U {\rm d}R - 
{\textstyle \frac{1}{2}} e^{-\eta} UV
{\rm d}w\, , \eqnum{\ref{bassae}b}
\addtocounter{equation}{1}
\end{eqnarray}
while as vector fields they are   
\begin{eqnarray}
k^{\mu} \partial/\partial x^{\mu} 
& = & e^{\eta} U^{-1}
\partial/\partial R \label{vectors} 
\eqnum{\ref{vectors}a}\\
l^{\mu} \partial/\partial x^{\mu} 
& = & e^{-\eta} \partial/\partial w - 
{\textstyle \frac{1}{2}} e^{-\eta} 
V \partial/\partial R - e^{-\eta} W^{a}
\partial/\partial x^{a} \eqnum{\ref{vectors}b}  
\, . 
\addtocounter{equation}{1}
\end{eqnarray}

Now define $u^\mu := \frac{1}{2} k^\mu + l^\mu$ and 
$n^\mu := \frac{1}{2} k^\mu - l^\mu$ along ${\cal N}$ as the timelike 
and spacelike unit normals of the $B$ two-surfaces. For each 
slice $B$ of the null hypersurface ${\cal N}$, the normals $u^\mu$ and $n^\mu$ 
determine a spanning spacelike three-surface $\Sigma$. As mentioned 
previously, the three-surface $\Sigma$ is not unique and, moreover, 
need not be defined throughout ${\cal M}$. Indeed, 
there is no guarantee that $u^\mu$ as defined is even surface-forming. 
(That is, in general $u_\mu$ does not satisfy the Fr\"obinius condition 
$u_{[\alpha}\nabla_{\mu} u_{\nu]}=0$.) Nevertheless, our construction 
provides us with what we need: a unit timelike vector $u^{\mu}$ orthogonal 
to $B$. We can therefore obtain an unreferenced 
energy surface density $k/\kappa$ which is the same for any 
slice or partial slice $\Sigma$ that 
contains $B$ and has timelike unit normal which agrees 
with $u^\mu$ at $B$.

Our construction implies 
\begin{equation}
u^{\mu}\partial/\partial x^{\mu} \rightarrow
\partial/\partial w
\end{equation}
on each ray as $R \rightarrow \infty$.
Now, the standard realization of the {\sc bms}-group Lie algebra (as 
vector fields on future null infinity) identifies the {\em extension} of
$\partial/\partial w$ to ${\cal I}^{+}$ (in a conformal completion
$\hat{\cal M}$ of ${\cal M}$) with a pure time translation.\cite{Sachs,review} 
Therefore, asymptotically, our fiducial surface $\Sigma_{\infty}$ determines 
precisely the pure time-translation generator of the {\sc bms} group.
We do {\em not} claim that $u^{\mu}$ generates an ``infinitesimal asymptotic 
symmetry transformation'' in the sense of Sachs\cite{Sachs}, 
{\it i.~e.~}that the various coefficients associated with the transformed metric 
$g_{\mu\nu} + 2\nabla_{(\mu} u_{\nu)}$ satisfy the fall-off conditions 
(\ref{Sachs}); however, this is unimportant for our construction.
 
\section{Computation of the quasilocal energy surface density}
We now turn to the task of calculating an expression for 
the unreferenced quasilocal energy surface density 
$\varepsilon |^{\scriptscriptstyle {\rm phy}} = k/\kappa$. 
Our starting point is the 
definition $k := - \sigma^{\mu\nu} \nabla_{\mu} n_{\nu}$, where the 
two-metric $\sigma^{\mu\nu} = g^{\mu\nu} + 2k^{(\mu} l^{\nu)}$ 
serves as the projection operator into $B$. We find it 
convenient to write\footnote{Note that the definition of $k$ does not
depend on how $n^{\mu}$ is extended off $B$.} $k = 2\mu + \rho$, 
where in the standard notation of the spin-coefficient 
formalism\cite{PenroseRindler} $- \mu$ and $\rho$ are, respectively, 
the expansions associated with the inward null normal and 
outward null normal to $B$. These are given by the formulae
\begin{eqnarray}
\mu & = & {\textstyle \frac{1}{2}} 
\sigma^{\mu\lambda} \nabla_{\mu} l_{\lambda}
= {\textstyle \frac{1}{2}}(\nabla_{\mu} l^{\mu} + k^{\nu} l^{\mu} 
\nabla_{\mu} l_{\nu})\label{spin} \eqnum{\ref{spin}a} \\
\rho & = & - {\textstyle \frac{1}{2}} 
\sigma^{\mu\lambda}\nabla_{\mu} k_{\lambda} = 
- {\textstyle \frac{1}{2}}(\nabla_{\mu} k^{\mu}
+ l^{\nu} k^{\mu} \nabla_{\mu} k_{\nu})\, .
\eqnum{\ref{spin}b}
\addtocounter{equation}{1}
\end{eqnarray} 
As a technical tool, it proves convenient to introduce
fiducial vector fields $\hat{k}^{\mu}$ and 
$\hat{l}^{\mu}$ determined from (\ref{vectors}) 
by setting $\eta = 0$ on $\cal N$. {}From the 
middle expressions above, it
is obvious that $\mu = e^{-\eta} \hat{\mu}$
and $\rho = e^{\eta} \hat{\rho}$, where the 
easier-to-calculate expressions $\hat{\mu}$ 
and $\hat{\rho}$ are built exactly as in 
(\ref{spin}) but with the fiducial 
null normals $\hat{k}^{\mu}$ and 
$\hat{l}^{\mu}$. Therefore, we may 
assume that $\eta = 0$ while 
calculating the spin coefficients in (\ref{spin}) and then simply 
multiply the $\eta = 0$ results by 
the appropriate factor to get the correct 
general expressions. Let us sketch the calculation. First, 
from (\ref{bassae}a) with
$\eta = 0$ note that $\hat{k}^{\mu} \nabla_{\mu} \hat{k}_{\nu} = 0$, 
because $\hat{k}_{\nu}$ is a gradient. Next, using both expressions
(\ref{bassae}) with $\eta = 0$, one 
can work the second term inside the parenthesis of 
(\ref{spin}a) into the form $\hat{k}^{\nu} 
\hat{l}^{\mu} \nabla_{\mu} \hat{l}_{\nu} = - U^{-1} \hat{l}^{\mu} 
\nabla_{\mu} U + \frac{1}{2} U \hat{k}^{\mu} \nabla_{\mu} V$. Finally, 
one writes the covariant-divergence terms as ordinary 
divergences; for example, $\nabla_{\mu} \hat{k}^{\mu} 
= (- g)^{-1/2} \partial_{\mu}(\sqrt{-g} 
\hat{k}^{\mu})$, where the square root of 
(minus) the determinant of the spacetime metric is 
$\sqrt{- g} = U\sqrt{\sigma}$. 

Following these steps and multiplying by 
the appropriate boost factors at the end of the calculation, one finds 
\begin{eqnarray}
\mu & = & {\textstyle \frac{1}{4}} 
e^{-\eta} \sigma^{-1} \dot{\sigma}
- {\textstyle \frac{1}{8}}
e^{-\eta} V \sigma^{-1}\sigma'
- {\textstyle \frac{1}{2}} e^{-\eta} 
\delta_{a} W^{a} \label{mu1} \eqnum{\ref{mu1}a} \\
\rho & = & - {\textstyle \frac{1}{4}} e^{\eta} U^{-1} 
\sigma^{-1}\sigma'{\,} . \eqnum{\ref{mu1}b}
\addtocounter{equation}{1}
\end{eqnarray}
Here the over-dot denotes partial differentiation 
by $\partial/\partial w$, the prime denotes partial 
differentiation by $\partial/\partial R$, and $\delta_{a}$ 
denotes the $B$ covariant 
derivative. Since $R$ is an areal radius, we may take 
$\sigma = R^{4}\sin^{2}\theta$ [see the form of the $B$
metric given in (\ref{Bmetric})]. Therefore, we obtain the 
compact expressions 
\begin{eqnarray}
\mu & = & - {\textstyle \frac{1}{2}} e^{-\eta} V R^{-1} 
- {\textstyle \frac{1}{2}} e^{-\eta} \delta_{a} W^{a} 
\label{mu2} \eqnum{\ref{mu2}a} \\
\rho & = & 
- e^{\eta} U^{-1} R^{-1}\, .
\eqnum{\ref{mu2}b}
\addtocounter{equation}{1}
\end{eqnarray} 
Adding twice (\ref{mu2}a) to (\ref{mu2}b), we arrive 
at our desired expression
\begin{equation}
k = - ( e^{-\eta} V + e^{\eta} U^{-1}) 
R^{-1} - e^{-\eta} \delta_{a} W^{a}\, ,
\end{equation}
which has the asymptotic form
\begin{equation}
k =  - 2R^{-1} + 2{\rm m}(w_{*},\theta,\phi) 
R^{-2} - \delta_{a} 
W^{a} + \Delta_{k}\, .
\label{kexpansion}
\end{equation}
Note that we have chosen not to expand the 
$O(R^{-2})$ pure divergence term $-\delta_{a}W^{a}$. Our assumption
about the fall-off of $\eta$ ensures that a term $-\eta^{2}/R$ which
appears in the asymptotic expression for $k$ can be 
swept into $\Delta_{k}$.

\section{The Bondi-Sachs mass}
Write the total quasilocal energy as 
$E = E |^{\scriptscriptstyle {\rm phy}} - 
E |^{\scriptscriptstyle {\rm ref}}$,
with the total {\em unreferenced} quasilocal 
energy $E |^{\scriptscriptstyle {\rm phy}}$ taken as
\begin{equation}
E |^{\scriptscriptstyle {\rm phy}} 
= \frac{1}{\kappa} \int_{B(w_{*},R)} 
{\rm d}^{2}x \sqrt{\sigma} k\, .
\label{unreferencedE1} \end{equation}
Plugging the expansion (\ref{kexpansion}) 
into the above expression, using
$\sqrt{\sigma} = R^{2}\sin\theta$ for our choice of coordinates, 
and integrating term-by-term, one finds 
\begin{equation}
E |^{\scriptscriptstyle {\rm phy}} = - R + M_{_{BS}}(w_{*}) 
+ \Delta_{E |^{\scriptscriptstyle {\rm phy}}}\, .
\label{unreferencedE2}
\end{equation}
Here the Bondi-Sachs mass associated with the $w = w_{*}$
cut of ${\cal I}^{+}$ is the two-surface
average of the mass aspect evaluated at $w = w_{*}$,\cite{Sachs,Goldberg}
\begin{equation}
M_{_{BS}}(w_{*}) 
= \frac{2}{\kappa} \int {\rm d}\Omega\, 
{\rm m}(w_{*},\theta,\phi)\, .
\end{equation}
We use the notation $\int {\rm d}\Omega := 
\int^{\pi}_{0} {\rm d}\theta
\int^{2\pi}_{0}{\rm d}\phi\sin\theta$ to denote proper
integration over the unit sphere 
(which is identified with a spherical cut of ${\cal I}^{+}$).
In passing from 
(\ref{unreferencedE1}) to (\ref{unreferencedE2}),
we have made an appeal to Stokes' theorem to show that 
the ``dangerous'' $O(R^{0})$ term that arises from proper 
integration over the pure-divergence term 
$\delta_{a} W^{a}$ in (\ref{kexpansion}) does indeed 
vanish. Hence, this term does not contribute to the 
Bondi-Sachs mass and does not spoil the result (\ref{unreferencedE2}).

The reference point contribution to the energy is
\begin{equation}
- E |^{\scriptscriptstyle {\rm ref}} = 
- \frac{1}{\kappa} \int_{B(w_{*},R)} 
{\rm d}^{2}x \sqrt{\sigma} 
k |^{\scriptscriptstyle {\rm ref}}\, ,
\label{referencedE1} 
\end{equation}
where the asymptotic expression for 
$k |^{\scriptscriptstyle {\rm ref}}$ must 
be determined from the specific asymptotic form (\ref{Bmetric}) of
the Sachs two-metric. We present this calculation in
the Appendix. The result is
\begin{equation}
- E |^{\scriptscriptstyle {\rm ref}} = 
R + 0 \cdot R^{0} + \Delta_{E |^{\scriptscriptstyle {\rm ref}}}\, .
\label{referencedE2}
\end{equation}
Note the absence of an $O(R^{0})$ term in 
$E |^{\scriptscriptstyle {\rm ref}}$. The result (3.5) 
has just the right form, in that it removes the part 
of $E |^{\scriptscriptstyle {\rm phy}}$ which becomes singular as 
$R \rightarrow \infty$ but does not itself contribute 
to the mass. Therefore the total quasilocal energy for large $R$ is 
\begin{equation}
E = \int_{B(w_{*},R)} 
{\rm d}^{2}x \sqrt{\sigma} \varepsilon 
= M_{_{BS}}(w_{*}) + \Delta_{E}\, .
\end{equation} 
This is the energy of the gravitational and matter fields associated with 
the spacelike three-surface $\Sigma$ which spans a $B$ slice 
of $\cal N$ and which tends toward $\Sigma_{\infty}$. 
Our main claim (\ref{limit}) follows immediately from (3.6). 

\section{Smeared Energy Surface Density}
Consider the expression $H_{B}$ for the on-shell value of the 
gravitational Hamiltonian appropriate for a spatially bounded 
three-manifold $\Sigma$, subject to the choice of a vanishing shift vector 
at the boundary $\partial\Sigma = B$:
\begin{equation}
  H_{B} = \int_{B}
  {\rm d}^{2}x \sqrt{\sigma} N \varepsilon
{\,} .
\label{boundaryH}
\end{equation} 
We refer to $H_B$ as the smeared energy surface density. 
Addition of this boundary term to the smeared Hamiltonian 
constraint ensures that as a whole the 
sum is functionally differentiable.\cite{BY} 
In this section we consider the $R\to\infty$ limit 
of $H_B$ along the null hypersurface $\cal N$ in exactly the same
fashion as we considered the limit (1.1) of the quasilocal energy previously. 
Before evaluating $\lim_{R \rightarrow \infty} H_{B(w_{*},R)}$,
let us discuss its physical significance. Consider a particular spherical 
cut $B(w_{*},\infty)$ of ${\cal I}^{+}$. A general {\sc bms} 
supertranslation pushes $B(w_{*},\infty)$ forward in retarded 
time $w$ in a general angle-dependent fashion. 
As is well-known, the infinitesimal generator corresponding 
to such a supertranslation has the form 
$\left. \alpha{\,} \partial/\partial w\right|_{{\cal I}^{+}}$, 
where $\alpha(\theta,\phi)$ is any twice differentiable 
function of the angular coordinates.\cite{Sachs} 
As we have seen, $\partial/\partial w$ is heuristically the hypersurface 
normal $u^{\mu}$ at $B(w_{*},\infty)$ of an asymptotic spanning 
three-surface $\Sigma_{\infty}$. In other words, each member
of the fleet of observers at $B(w_{*},\infty)$ rides along
$\partial/\partial w$. Therefore, again heuristically, the 
on-shell value of the Hamiltonian generator of a general {\sc bms} 
supertranslation is 
\begin{equation}
\int_{B(w_{*},\infty)}{\rm d}^{2}x
\sqrt{\sigma} \alpha \varepsilon\, .
\end{equation} 
This symbolic expression coincides with the $R\to\infty$ limit 
of the smeared energy surface density (4.1), where we
set $\alpha(\theta,\phi)
:= \lim_{R \rightarrow \infty} N(R,\theta,\phi)$ (suitable
fall-off behavior for $N$ is assumed). Thus, 
$\lim_{R \rightarrow \infty} H_{B(w_{*},R)}$ 
defines a physically meaningful element in the dual space 
of general supertranslations. In this respect it is like the 
``supermomentum" of Geroch.\cite{Geroch} In the next section we show 
explicitly that, in fact, $\lim_{R \rightarrow \infty} 
H_{B(w_{*},R)}$ is precisely Geroch's ``supermomentum." Note, 
however, that it might be 
better to call such an expression the ``superenergy,'' as it arises 
entirely from the ``energy sector'' of the Hamiltonian's boundary term  
(that is, the sector with vanishing shift vector) but also incorporates 
the ``many-fingered'' nature of time (that is, an arbitrary lapse function). 

Let us now evaluate the $R\to\infty$ limit of the smeared energy surface
density $H_B$. 
As we have stated, there is no $O(R^{0})$ contribution to 
$E |^{\scriptscriptstyle {\rm ref}}$. 
The absence of this contribution stems from 
the fact that the two-sphere average of the coefficient 
${}^{(2)}\!k |^{\scriptscriptstyle {\rm ref}}$ of the 
$O(R^{-2})$ piece of the 
reference term $k |^{\scriptscriptstyle {\rm ref}}$ vanishes. 
As spelled out in the Appendix, this fact follows directly 
from an equation governing the required isometric 
embedding of $B$ into Euclidean three-space. Moreover, as  seen in 
Section 2, the coefficient  ${}^{(2)}\! k$ of the $O(R^{-2})$ piece of the 
physical $k$ is 
not solely twice the mass aspect but also contains a unit-sphere 
divergence term. Now, in the present case $\varepsilon 
= (k - k |^{\scriptscriptstyle {\rm ref}})/\kappa$ is smeared 
against a function 
$N$, so one might worry that the limit is spoiled in some way by
the presence of the smearing function. However, as we now show,
{\em for solutions of the field equations}, the {\em unintegrated} 
expression $4\pi R^{2}\varepsilon$ is precisely the mass aspect 
of the system in the $R \rightarrow \infty$ limit. This striking 
result rests on an exact cancellation between 
${}^{(2)}\!k |^{\scriptscriptstyle {\rm ref}}$ and the aforementioned 
unit-sphere divergence part of ${}^{(2)}\!k$. 

With the machinery set up in the previous sections and the Appendix 
[see in particular equations (\ref{kexpansion}) and (A2)], we find 
the following limit:
\begin{equation}
\lim_{R \rightarrow \infty} 
{\textstyle \frac{1}{2}}\kappa R^{2} \varepsilon  = 
{\rm m} (w_{*},\theta,\phi) -
{\textstyle \frac{1}{2}}
\left[\csc\theta\, \partial_{a}
(\sin\theta {}^{(2)}\! W^{a})  
-{\textstyle \frac{1}{2}}
{}^{(3)}\!{\cal R} \right] \, .
\label{limk-k0}
\end{equation}
Here we set $\kappa = 8\pi$ (in geometrical units) and use 
${}^{(3)}\!{\cal R}$ to denote the coefficient of the $O(R^{-3})$ 
piece of the $B$
Ricci scalar. Also, the coefficients ${}^{(2)}\! W^{a}$ of the leading
$O(R^{-2})$ pieces of $W^{a}$ are listed in (1.2c). 
Inspection of (\ref{Sachs}c,d) shows that ${}^{(2)}\!W^{a}$ 
is expressed in terms of the same functions, $X$ and $Y$, that
appear in the $O(R^{-1})$ piece of $\sigma_{ab}/R^{2}$. Furthermore, 
a short calculation with the $B$ metric shows that 
for these solutions ${}^{(3)}\! {\cal R}$ may be 
expressed in terms of ${}^{(2)}\! W^{a}$ as follows:
\begin{equation}
- {\textstyle \frac{1}{2}} {}^{(3)}\! {\cal R} = 
- \csc\theta\, \partial_{a}(\sin\theta
{}^{(2)}\! W^{a}) \, .
\label{calR=dW}
\end{equation}
Therefore, the term in (\ref{limk-k0}) which is enclosed by square brackets
vanishes, and we obtain
\begin{equation}
\lim_{R \rightarrow \infty} 
\int_{B(w_{*},R)} {\rm d}^{2}x \sqrt{\sigma} N
\varepsilon =
\frac{2}{\kappa}\int {\rm d}\Omega{\,} \alpha(\theta,\phi) {\rm m}
(w_{*},\theta,\phi)
{\,} , \label{limitresult}
\end{equation}
for the desired limit. 
This result shows that the $R\to\infty$ limit of the smeared energy surface 
density equals the smeared mass aspect. Coupled with the findings 
of the next section, it 
follows that Geroch's ``supermomentum" is just the smeared mass aspect. 
This simple result does not appear to be widely known. 

Finally, recall that the Bondi-Sachs 
four-momentum components\footnote{Underlined Greek indices
refer to components of the total Bondi-Sachs four-momentum.} 
$P_{_{BS}}^{\underline{\lambda}}$ correspond asymptotically to a pure 
translation. In terms of the smeared energy surface density, one 
obtains a pure translation for a judicious choice of lapse function  
on $B(w_{*},\infty)$; namely, $\alpha(\theta,\phi) 
= \epsilon_{\underline{\lambda}}
\alpha^{\underline{\lambda}}(\theta,\phi)$, where the
$\epsilon_{\underline{\lambda}}$ are constants and \cite{Goldberg}
\begin{eqnarray}
\alpha^{\underline{0}} & = & 1
\label{translations} \eqnum{\ref{translations}a} \\
\alpha^{\underline{1}} & = & \sin\theta\cos\phi 
\eqnum{\ref{translations}b} \\
\alpha^{\underline{2}} & = & \sin\theta\sin\phi 
\eqnum{\ref{translations}c} \\
\alpha^{\underline{3}} & = & \cos\theta
\eqnum{\ref{translations}d}\, .
\addtocounter{equation}{1}
\end{eqnarray}
Therefore, we write $\epsilon_{\underline{\lambda}} 
P_{_{BS}}^{\underline{\lambda}}(w_{*}) = 
\lim_{R \rightarrow \infty} H_{B(w_{*},R)}$ for the appropriate limiting
value of $N$, and thereby obtain the Bondi-Sachs four-momentum as a 
Hamiltonian value. 

\section{Supermomentum} 
In this section we show that the null limit of the smeared Hamiltonian 
boundary value, Eq.~(4.1), is the ``supermomentum" of Geroch.\cite{Geroch} 
To be precise, we show that in the null limit $H_{B}$
equals Geroch's ``supermomentum" as written by Dray and 
Streubel.\cite{DrayStreubel} 
The spin-coefficient formalism is required for this 
analysis.\footnote{Throughout $\S$V we deal 
exclusively with smooth expansions in inverse powers of an 
{\em affine} radius, as we know of no work examining the standard spin 
coefficient approach to null infinity within a more general 
framework such as the polyhomogeneous one. 
The expansions we borrow from \cite{Dougan} 
are valid for Einstein-Maxwell theory.} 
Apart from a few minor notational changes we adopt the 
conventions of Dougan.\cite{Dougan} 
Geometrically, the scenario is nearly the same as the one described 
in the previous sections. However, we now work with a 
slightly different type of Bondi coordinates. Namely, 
$(w,r,\zeta,\bar{\zeta})$, where $r$ is an {\em affine} parameter 
along the null-geodesic generators of ${\cal N}$ and $\zeta = 
e^{{\rm i}\phi}\cot(\theta/2)$ is the stereographic coordinate. 
Dougan picks\footnote{Our $k_{\mu}$ and $l_{\mu}$ 
respectively correspond to 
$l_{a} = \nabla_{a} u$ and $n_{a}$ in \cite{Dougan}, 
where $u$ is Dougan's
retarded time. The minus sign difference
between our definition for $k_{\mu}$ and Dougan's 
definition for $l_{a}$ stems
from a difference in metric-signature conventions 
[ours is $(-,+,+,+)$]. The convention for
metric signature does not affect the spin coefficients
(\ref{spinexpansions}).}
$k_{\mu} = - \nabla_{\mu} w$ as the first leg of a null tetrad, 
which is the same normal as given in (\ref{bassae}a) if $\eta = 0$. 
For convenience, in this section we ignore the kinematical freedom
associated with the $\eta$ parameter, setting it to zero throughout. 
As before, the vector field $u^{\mu} := \frac{1}{2} k^{\mu} + 
l^{\mu}$ (equality restricted to ${\cal N}$) defines a three-surface 
$\Sigma$ spanning each $B$ slice of ${\cal N}$. 
It follows that $u^{\mu}\partial/\partial x^{\mu} 
\rightarrow \partial/\partial w$
as $r \rightarrow \infty$, and hence our asymptotic slice 
$\Sigma_{\infty}$ again defines a pure {\sc bms} time translation.

Let us first collect the essential background results from 
\cite{Dougan} which we will need. First, the required spin 
coefficients have the following asymptotic expansions:\footnote{Note 
the dual use of $\sigma$ as both the stem letter for
the $B$ two-metric and as the spin coefficient known as the shear. 
We have used $\sigma$ twice in order to stick with the 
conventions of our references as much as possible. In all but 
equation (\ref{spinexpansions}), where $\sigma$ has the spin-coefficient 
meaning, it carries a ``$0$'' superscript denoting the asymptotic piece.}
\begin{eqnarray}
  \rho & = & - r^{-1} - \sigma^{0} 
  \bar{\sigma}{}^{0} r^{-3} + O(r^{-5})
\nonumber \\
\mu & = & 
- {\textstyle \frac{1}{2}} r^{-1} 
- [\Psi^{0}_{2} + \sigma^{0} 
  \dot{\bar{\sigma}}{}^{0}
+ \ed^{2}_{\naught} 
  \bar{\sigma}{}^{0}]r^{-2} 
+ O(r^{-3})
\nonumber \\
  \sigma & = & 
  \sigma^{0} r^{-2} 
+ O(r^{-4})
\label{spinexpansions}
{\,} ,
\end{eqnarray}
where $\mu$ is Dougan's $- \rho'$, the term $\Psi^{0}_{2}$ is a certain 
asymptotic component of the Weyl tensor, and $\sigma^{0}$ 
is the asymptotic piece of the shear. Like before, an over-dot denotes 
differentiation by $\partial/\partial w$. As fully described in 
\cite{Dougan}, $\ednaught$ is the standard differential operator from 
the compacted spin-coefficient formalism, here defined on the {\em unit} 
sphere. The expansion for the corresponding full operator in spacetime 
is
\begin{equation}
  \ed = r^{-1} \ednaught 
+ r^{-2} [s(\edbarnaught\sigma^{0}) 
- \sigma^{0} \edbarnaught] + O(r^{-3})
{\,} ,
\end{equation}
where $s =$ {\sc sw}$(\varphi)$, $\varphi$ being the spacetime 
scalar on which $\ed$ acts and {\sc sw} denoting {\em spin weight}. 
The commutator of $\ed$ and $\bar{\ed}$ is
\begin{equation}
(\bar{\ed}\ed - \ed \bar{\ed})\varphi = {\textstyle \frac{1}{2}}
s {\cal R} \varphi {\,} .
\label{commutator}
\end{equation}
Now consider the following ansatz for the $B$ intrinsic 
Ricci scalar:
\begin{equation}
{\cal R} = 2r^{-2} + {}^{(3)}\!{\cal R} r^{-3} + O(r^{-4})
{\,} .
\end{equation}
If we insert this expansion into (\ref{commutator}) and
expand both sides of the equation [assuming $\varphi =
{}^{(0)}\!\varphi + {}^{(1)}\!\varphi r^{-1} + O(r^{-2})$ 
with {\sc sw}$(\varphi)
= 1$], then to lowest order, namely $O(r^{-2})$, we get a trivial
equality. However, equality at the next order demands that
\begin{equation}
{\textstyle \frac{1}{2}} {}^{(3)}\!{\cal R} 
= \edbarnaught^{2} \sigma^{0}
+ \ednaught^{2} \bar{\sigma}{}^{0}
{\,} .
\label{result}
\end{equation}
This will prove to be a very important result for our purposes.
Finally, Dougan gives the following expansion for the $B$ volume element:
\begin{equation}
  {\rm d}^{2}x\sqrt{\sigma} = {\rm d}\Omega {\,} r^{2} 
  (1 - \sigma^{0} \bar{\sigma}^{0} r^{-2}) 
+ O(r^{-2})
{\, } .
\end{equation} 
(Here $\sigma$ is the determinant of the $B$ metric and $\sigma^0$ 
is the asymptotic piece of the shear.) 

We now consider the spin-coefficient expression for the smeared energy
surface density introduced in $\S$IV. Again with $k = 2\mu + \rho$, 
in the present notation we find
\begin{equation}
k =  - 2r^{-1} - 
          2[\Psi^{0}_{2} + \sigma^{0} \dot{\bar{\sigma}}{}^{0}
           + \ed^{2}_{\naught} 
\bar{\sigma}{}^{0}]r^{-2} + O(r^{-3})
{\,} .
\end{equation}
Moreover, by an argument identical to the one found in the last 
paragraph of the Appendix (although here with the affine 
radius $r$ rather than the areal radius $R$), 
we know that the result (\ref{result}) determines
\begin{equation}
k |^{\scriptscriptstyle {\rm ref}} = - 2 r^{-1}
-(\edbarnaught^{2} \sigma^{0}
+ \ednaught^{2} \bar{\sigma}{}^{0}) r^{-2} + O(r^{-3})
\end{equation}
as the appropriate asymptotic expansion for the reference term.
Therefore, ($\kappa$ times) the full quasilocal energy surface density is
\begin{equation}
\kappa\varepsilon = 
           - 2[\Psi^{0}_{2} + \sigma^{0} \dot{\bar{\sigma}}{}^{0}
           + {\textstyle \frac{1}{2}}\ednaught^{2} 
\bar{\sigma}{}^{0} - {\textstyle \frac{1}{2}}\edbarnaught^{2} \sigma^{0}
]r^{-2} + O(r^{-3})
{\,} .
\end{equation}
At this point we consider again 
a smearing function $N$, with appropriate 
fall-off behavior and limit $\alpha(\zeta,\bar{\zeta}) 
= \lim_{r \rightarrow \infty} N(r,\zeta,\bar{\zeta})$. Using the results
amassed up to now, one computes that the limit of the smeared energy
surface density is
\begin{equation}
\lim_{r \rightarrow \infty} \int_{B(w_{*}, r)}
{\rm d}^{2}x \sqrt{\sigma} N\varepsilon = - \frac{2}{\kappa} \int
{\rm d}\Omega{\,}\alpha\left. [\Psi^{0}_{2} 
+ \sigma^{0} \dot{\bar{\sigma}}{}^{0}
           + {\textstyle \frac{1}{2}}\ednaught^{2} 
\bar{\sigma}{}^{0} - 
{\textstyle \frac{1}{2}}\edbarnaught^{2} \sigma^{0}]\right|_{w = w_{*}}
{\,} .
\label{superE}
\end{equation}
The right-hand side of this equation is the ``supermomentum'' of Geroch as 
written by Dray and Streubel [see equation (A1.12) of \cite{DrayStreubel}
and set their $b = 0$ for a Bondi frame as we have here]; and the ``supermomentum''
is known to be the ``charge integral" associated with the 
Ashtekar-Streubel flux\cite{AshtekarStreubel}
of gravitational radiation at ${\cal I}^{+}$ (in the restricted case 
when the flux is associated with a supertranslation).\cite{Shaw}
Dray and Streubel
have discussed the importance of the particular factors of $\frac{1}{2}$ 
which multiply the last two terms within the square brackets on the 
right-hand side of equation (\ref{superE}). It is evident from our approach
that the origin of these $\frac{1}{2}$ factors stems from the flat-space 
reference of the quasilocal energy (flat-space being the correct reference 
in the present context). 
When $\alpha$ determines 
a pure {\sc bms} translation, the last two terms in the integrand integrate 
to zero. For instance, setting $\alpha = 1$, one finds that the strict energy
\begin{equation}
E = \int_{B(w_{*},\infty)} 
{\rm d}^{2}x\sqrt{\sigma}
\varepsilon = - \frac{2}{\kappa}
\int {\rm d}\Omega \left. [\Psi^{0}_{2} + \sigma^{0} 
\dot{\bar{\sigma}}{}^{0}]\right|_{w = w_{*}}
{\,} 
\end{equation}
is the standard spin-coefficient expression for the Bondi-Sachs mass 
$M_{_{BS}}(w_{*})$.\cite{Dougan,ExtonNewmanPenrose}
 
\section*{Acknowledgments}
We thank H. Balasin, P. T. Chru\'{s}ciel, T. Dray, and N. \'{O} Murchadha
for helpful discussions and correspondence. We acknowledge support from National
Science Foundation grant 94-13207. S.\ R.\ Lau has been chiefly supported by 
the ``Fonds zur F\"{o}r\-der\-ung der wis\-sen\-schaft\-lich\-en For\-schung'' 
in Austria (FWF project P 10.221-PHY and Lise Meitner Fellowship M-00182-PHY).

\appendix
\section*{Subtraction term}
In this Appendix we prove that 
the subtraction-term contribution
$- E |^{\scriptscriptstyle {\rm ref}}$ to the quasilocal energy obeys
\begin{equation}
- E |^{\scriptscriptstyle {\rm ref}} = 
R + 0 \cdot R^{0} + 
\Delta_{E |^{\scriptscriptstyle {\rm ref}}}  \label{mainresult}
\end{equation}
for large $R$. Moreover, we derive the result 
\begin{equation} 
{}^{(2)}\! k |^{\scriptscriptstyle {\rm ref}} = 
-{\textstyle\frac{1}{2}} {}^{(3)}\!{\cal R} 
\label{2k0=3calR} 
\end{equation} 
relating the coefficient ${}^{(2)}\! k |^{\scriptscriptstyle {\rm ref}}$ of the
$O(R^{-2})$  piece of the reference term 
$k |^{\scriptscriptstyle {\rm ref}}$ and the coefficient ${}^{(3)}\!{\cal R}$ 
of the $O(R^{-3})$ piece of the $B$ Ricci scalar ${\cal R}$. 

The $- E |^{\scriptscriptstyle {\rm ref}}$ term is constructed as 
follows. Consider a generic $B$ slice of $\cal N$. 
Assume that $B$ may be embedded isometrically in 
Euclidean three-space $E^{3}$ and that the embedding 
is suitably unique (we address these issues below). 
Let $(k |^{\scriptscriptstyle {\rm ref}})_{ab}$ 
represent the extrinsic curvature 
of $B$ as isometrically embedded in $E^{3}$. 
The flat-space reference density is 
$\varepsilon |^{\scriptscriptstyle {\rm ref}} =
k |^{\scriptscriptstyle {\rm ref}}/\kappa$, and in 
terms of this density 
\begin{equation}
- E |^{\scriptscriptstyle {\rm ref}} = - \int_{B(w_{*},R)} 
{\rm d}^{2}x \sqrt{\sigma} 
\varepsilon |^{\scriptscriptstyle {\rm ref}}\, . \label{integral}
\end{equation}
Since it is the intrinsic geometry of $B$ which determines 
the reference term $k |^{\scriptscriptstyle {\rm ref}}$, 
let us first collect a few 
results concerning this geometry. In the Bondi coordinate 
system the two-metric 
of $B$ takes the following form:\cite{Sachs}
\begin{eqnarray}
\lefteqn{\sigma_{ab} {\rm d}x^{a} {\rm d}x^{b} =}
& & \nonumber \\
& & {\textstyle \frac{1}{2}} R^{2} 
(e^{2\gamma} + e^{2\delta}){\rm d}\theta^{2}
+ 2 R^{2} \sin\theta\sinh(\gamma - \delta){\rm d}\theta{\rm d}\phi
+ {\textstyle \frac{1}{2}} R^{2} 
(e^{-2\gamma} + e^{-2\delta})\sin^{2}\theta{\rm d}\phi^{2}\, .
\label{Bmetric} \end{eqnarray}
In terms of the functions $X$ and $Y$, $\gamma$ and $\delta$ 
have the expansions
[see (\ref{Sachs}d)]
\begin{eqnarray}
\gamma & = & (X + Y)
R^{-1} + \Delta_{\gamma} \nonumber \\
& & \\
\delta & = & (X - Y)
R^{-1} + \Delta_{\delta}\, . 
\label{moreBmetric}\nonumber 
\end{eqnarray}
As mentioned, we do not necessarily enforce Sachs' ``outgoing radiation
condition'' which would, in fact, imply that in the expansions for 
$\gamma$ and $\delta$ the terms that appear after the leading $O(R^{-1})$ 
terms are $O(R^{-3})$.\cite{Sachs} 

{}From the form of the line element (\ref{Bmetric}) one easily
verifies (\ref{Sachs}d) and that $\sqrt{\sigma} = R^{2}\sin\theta$. 
A bit more work establishes that the scalar curvature 
${\cal R}$ of $B$ has the asymptotic form 
\begin{equation}
{\cal R} = 2R^{-2} +\,\! ^{(3)}\! {\cal R} R^{-3} 
+ \Delta_{{\cal R}}\, ,
\label{curvexpand}
\end{equation}
where $^{(3)}\! {\cal R} 
= \,\! ^{(3)}\! {\cal R}(w_{*},\theta,\phi)$. We have given the explicit 
form (\ref{calR=dW}) of the coefficient $^{(3)}\! {\cal R}$ corresponding 
to the asymptotic solutions considered here, and we have found this 
coefficient to be a pure divergence on the unit sphere. That this term 
integrates to zero can also be shown {\em via} the argument to follow, 
which does not assume that the Einstein equations hold.

\noindent \underline{\em Lemma:} The two-sphere average 
of ${}^{(3)}\! {\cal R}$ vanishes, {\em i.~e.}
$\int {\rm d}\Omega\,{}^{(3)}\! {\cal R}
= 0$. 
To prove the lemma, start with the Gauss-Bonnet 
theorem \cite{O'Neill} 
\begin{eqnarray}
8\pi & = & \int_{B}{\rm d}^{2}x \sqrt{\sigma} {\cal R} \nonumber \\
     & = & \int_{B}{\rm d}^{2}x \sin\theta \left(2 +\,\! 
^{(3)}\! {\cal R} R^{-1} + R^{2}\Delta_{{\cal R}}\right) \, . 
\end{eqnarray}
In the second line we have simply expanded $\cal R$ and used 
$\sqrt{\sigma} = R^{2}\sin\theta$. On the right side of the 
equation, integration over the first term inside the brackets 
gives $8\pi$. Therefore, we arrive at
\begin{equation}
0 = \int {\rm d}\Omega{\,} {}^{(3)}\! {\cal R}
+ R^{3}\int {\rm d}\Omega{\,}\Delta_{{\cal R}}\, .
\end{equation}
Since $\int {\rm d}\Omega{\,}\Delta_{{\cal R}}$ falls off 
{\em faster} than $O(R^{-3})$,
the $R \to \infty$ limit of this last equation proves the lemma. 

Now let us discern the $R$ dependence of the reference term
$k |^{\scriptscriptstyle {\rm ref}}$. We must first address 
the issue of whether or not 
$B$ may be isometrically embedded in $E^{3}$ in a suitably 
unique way. It is known that a Riemannian manifold possessing 
two-sphere topology
and everywhere positive scalar curvature can be globally immersed
in $E^{3}$. An immersion differs from an embedding by allowing for
self-intersection of the surface (seemingly allowable provided that
$k |^{\scriptscriptstyle {\rm ref}}$ remains well defined). 
The Cohn-Vossen theorem states 
that any compact two-surface contained in $E^{3}$ whose curvature is 
everywhere positive is unwarpable. Unwarpable means that the 
surface is uniquely determined by its two-metric, up to 
translations or rotations in $E^{3}$ (which of course do not
affect $k|^{\scriptscriptstyle {\rm ref}}$).\cite{BY,subtraction} 
Our surface $B$ has
two-sphere topology, and for $R$ sufficiently large the scalar 
curvature $\cal R$ is everywhere positive. It thus follows that 
for suitably large $R$, the reference term 
$k |^{\scriptscriptstyle {\rm ref}}$ is well defined. 

The key equation for determining the form of 
$k |^{\scriptscriptstyle {\rm ref}}$ is the 
standard Gauss-Codazzi relation:\cite{O'Neill}
\begin{equation}
(k |^{\scriptscriptstyle {\rm ref}})^{2} - 
(k |^{\scriptscriptstyle {\rm ref}})^{a}_{b}
(k |^{\scriptscriptstyle {\rm ref}})^{b}_{a}
- {\cal R} = 0\, . \label{key}
\end{equation}
This equation (essentially a two-dimensional version of the
Hamiltonian constraint) is an integrability criterion, 
obeyed by our embedding, which relates certain components of 
the {\em vanishing} Riemann 
tensor of $E^{3}$ to the intrinsic $B$ curvature scalar 
and the desired reference extrinsic curvature
tensor $(k |^{\scriptscriptstyle {\rm ref}})_{ab}$. 
Next, since $B$ approaches a perfectly
round sphere as $R \rightarrow \infty$, we take the following 
expansions for the various pieces of 
$(k |^{\scriptscriptstyle {\rm ref}})^{a}_{b}$ as an ansatz:
\begin{eqnarray}
(k |^{\scriptscriptstyle {\rm ref}})^{\theta}_{\theta} & = & - R^{-1} 
+\,\! ^{(2)}\! 
(k |^{\scriptscriptstyle {\rm ref}})^{\theta}_{\theta} R^{-2} 
+ \Delta_{(k)}
\label{expand} \eqnum{\ref{expand}a} \\
(k |^{\scriptscriptstyle {\rm ref}})^{\phi}_{\phi} & = & - R^{-1} 
+\,\! ^{(2)}\! (k |^{\scriptscriptstyle {\rm ref}})^{\phi}_{\phi} R^{-2} 
+ \Delta_{(k)}
\eqnum{\ref{expand}b} \\
(k |^{\scriptscriptstyle {\rm ref}})^{\theta}_{\phi} & = & 0 \cdot R^{-1} 
+ \Delta_{(k)}
\eqnum{\ref{expand}c} \\
(k |^{\scriptscriptstyle {\rm ref}})^{\phi}_{\theta} 
& = & 0 \cdot R^{-1} + \Delta_{(k)} {\,} .
\eqnum{\ref{expand}d} 
\addtocounter{equation}{1}
\end{eqnarray}
To avoid clutter, for the various remainder terms we have 
used in (\ref{expand}) 
simply a subscript $(k)$ in place of what should be 
$(k |^{\scriptscriptstyle {\rm ref}})^{\theta}_{\theta}$, 
{\em etc}. Plugging the expansions (\ref{curvexpand}) and 
(\ref{expand}) into (\ref{key}), one finds that 
to lowest order, namely $O(R^{-2})$, the equation is identically satisfied. 
At the next order, namely $O(R^{-3})$, the equation (\ref{key}) yields 
the result (A2).
Therefore, we have found the following asymptotic expansion
for $k |^{\scriptscriptstyle {\rm ref}}$:
\begin{equation}
k |^{\scriptscriptstyle {\rm ref}} = - 2R^{-1} - 
{\textstyle \frac{1}{2}}\,\!
^{(3)}\! {\cal R} R^{-2} + \Delta_{k |^{\scriptscriptstyle {\rm ref}}}\, .
\label{k0expansion}
\end{equation} 
The first lemma we proved above has an important consequence. 
It ensures that the ``dangerous'' $O(R^{0})$ term in the
integral (\ref{integral}) vanishes (regardless of whether or not
(\ref{calR=dW}) holds, which requires that the Einstein equations 
are satisfied asymptotically). Hence we get the
result (\ref{mainresult}).


\begin{figure}
\epsfxsize=6in
\centerline{\epsfbox{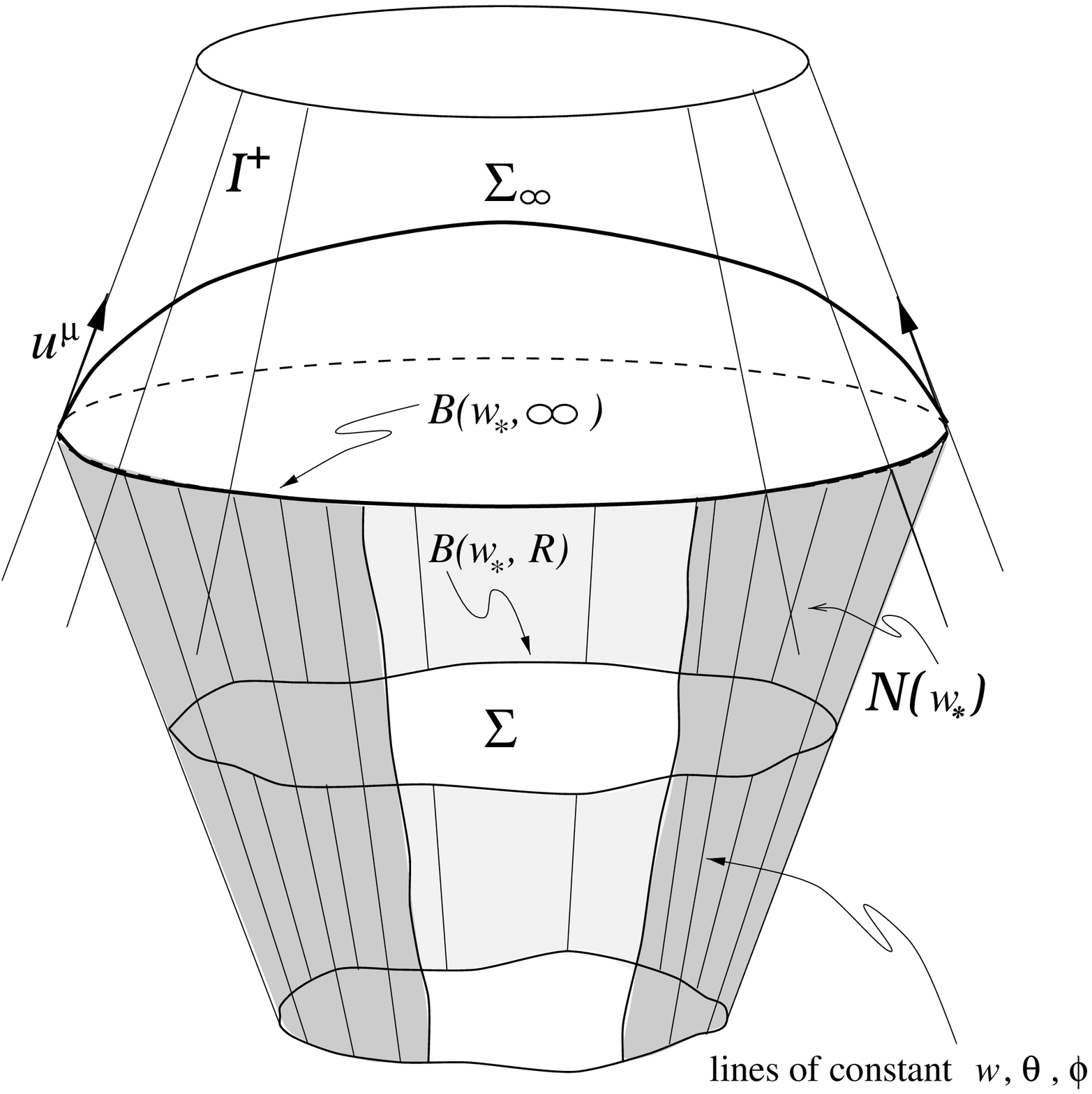}}
\caption{Geometry of the asymptotic slice $\Sigma_{\infty}$}
{\small 
In this figure one dimension of the two-surfaces 
$B(w_{*},R)$ is suppressed. The shaded, partially cut-away, 
conical surface depicts the null hypersurface 
${\cal N} = {\cal N}(w_{*})$ determined by a
constant value $w = w_{*}$ of retarded time.
Heuristically, in the limit $R \rightarrow \infty$ the $\Sigma$ 
slice spanning $B(w_{*}, R)$ becomes the 
asymptotic slice $\Sigma_{\infty}$ which spans a round
spherical cut of ${\cal I}^{+}$, and one should envision the 
spacelike slice $\Sigma_{\infty}$ as becoming null 
asymptotically. Although $\partial/\partial w$ is timelike 
everywhere in the {\em physical} spacetime
${\cal M}$ (at least in relevant exterior regions), 
the {\em extension} to ${\cal I}^{+}$ 
of $\partial/\partial w$ (in
a conformal completion $\hat{\cal M}$ of ${\cal M}$) 
is a null vector which
lies in ${\cal I}^{+}$. Again heuristically, on $B(w_{*},\infty)$ the
$\Sigma_{\infty}$ hypersurface normal 
$u^{\mu}\partial/\partial x^{\mu}$ is $\partial/\partial w$.}  
\end{figure} 

\end{document}